\documentclass[12pt]{article}
\usepackage{graphicx}
\usepackage{amsmath}
\usepackage{amssymb}
\usepackage{caption2}
\begin{document}
\bibliographystyle{prsty}
\begin{center}
{\large {\bf \sc{   Radiative decays of the  $D_{s0}(2317)$,
$D_{s1}(2460)$ and the related strong coupling constants  }}} \\[2mm]
Zhi-Gang Wang \footnote{E-mail,wangzgyiti@yahoo.com.cn.  }    \\
Department of Physics, North China Electric Power University,
Baoding 071003, P. R. China
\end{center}

\begin{abstract}
In this article, we take the point of view that the  charmed
 mesons $D_{s0}(2317)$ and  $D_{s1}(2460)$ with the spin-parity $0^+$ and $1^+$ respectively are
  the conventional $c\bar{s}$
states, and  calculate the strong coupling constants $G_S$( for
$\langle D_s^* \phi | D_{s0} \rangle$ ) and $G_A$( for $\langle D_s
\phi | D_{s1} \rangle$ ) in the framework of the light-cone QCD sum
rules approach. The  strong coupling constants $G_S$ and $G_A$ are
related to the basic parameter $\hat{\mu}$ in the heavy quark
effective Lagrangian, the numerical value is larger than the
existing estimation. With the assumption of the vector meson
dominance of the intermediate $\phi(1020)$, we study the radiative
decays $D_{s0}\to D_s^* \gamma $ and $D_{s1}\to D_s \gamma $.
\end{abstract}


\section{Introduction}
 The observation of the two charmed resonances
$D_{s0}(2317)$ in the  $D_s\pi^0$ invariant mass distribution and
$D_{s1}(2460)$  in the $D_s^* \pi^0$ and $D_s\gamma$ mass
distributions has triggered hot debate on their nature,
under-structures and whether it is necessary to introduce the exotic
states \cite{exp03,Swanson06}. They can not be comfortably
identified as the quark-antiquark bound states in the spectrum of
the constituent quark models, their masses  are significantly lower
than the values of the   $0^+$ and $1^+$ states  respectively from
the quark models and lattice simulations \cite{QuarkLattice}. The
difficulties to identify the $D_{s0}(2317)$ and $D_{s1}(2460)$
states with the conventional $c\overline{s}$ mesons are rather
similar to those appearing in the light scalar mesons below $1GeV$.
The light scalar mesons are the subject of an intense and continuous
controversy in clarifying the hadron spectroscopy \cite{Godfray},
the more elusive things are the constituent structures of the
$f_0(980)$ and $a_0(980)$ mesons with almost the degenerate masses.
The mesons $D_{s0} (2317)$ and $D_{s1} (2460)$ lie just below the $D
K$ and $D^\ast K$ threshold respectively,  which are analogous to
the situation that the scalar mesons  $a_0(980)$ and $f_0(980)$ lie
just below the $K\bar{K}$ threshold and  couple strongly to the
nearby channels. The mechanism responsible for the low-mass charmed
mesons may be the same as the light scalar nonet mesons, the
$f_{0}(600)$, $f_{0}(980)$, $a_{0}(980)$  and $K^{\ast}_{0}(800)$
\cite{ColangeloWang,WangWan06,Wang0611,ReviewScalar,WangScalar05}.
There have been a lot of explanations for their nature, for example,
the conventional $c\bar{s}$ states \cite{2quark}, two-meson
molecular states \cite{2meson}, four-quark states \cite{4quark},
etc\footnote{The  literatures listed here are far from complete, for
more literatures, one can consult Ref.\cite{Swanson06}. }. If we
take the scalar mesons $a_0(980)$ and $f_0(980)$ as four quark
states with the
 constituents  of scalar diquark-antidiquark  sub-structures, the
masses of the scalar nonet mesons below  $1GeV$ can be naturally
explained \cite{ReviewScalar,WangScalar05}.

There are other possibilities besides the four-quark state
explanations, for example, the scalar mesons $a_0(980)$, $f_0(980)$,
$D_{s0}(2317)$ and the axial-vector meson $D_{s1}(2460)$ may have
bare $P-$wave $q\overline{q}$ and $c\bar{s}$ kernels   with strong
coupling to the nearby thresholds respectively, the $S-$wave virtual
intermediate hadronic states (or the virtual mesons loops) play a
crucial role in the composition of those bound states (or resonances
due to the masses below or above the thresholds). The hadronic
dressing mechanism (or unitarized quark models) takes the point of
view that the mesons $f_0(980)$, $a_0(980)$, $D_{s0}(2317)$ and
$D_{s1}(2460)$ have small $q\bar{q}$ and $c\bar{s}$  kernels of the
typical $q\bar{q}$ and $c\bar{s}$  mesons size respectively. The
strong couplings to the virtual intermediate hadronic states (or the
virtual mesons loops) may result in smaller masses than the
conventional scalar $q\bar{q}$   and $c\bar{s}$  mesons in the
constituent quark models, enrich the pure $q\bar{q}$ and $c\bar{s}$
states with other components \cite{HDress,UQM}. Those mesons may
spend part (or most part) of their lifetime as virtual $ K \bar{K}
$, $DK$ and $D^*K$ states
\cite{ColangeloWang,WangWan06,Wang0611,HDress,UQM}.

The radiative decays  can be used to probe the under-structures of
the hadrons, and  they are suitable  to understand  the nature of
the $D_{s0}(2317)$, $D_{s1}(2460)$ and distinguish among different
interpretations
\cite{Colangelo1,Colangelo2,Bardeen3,Godfrey4,Li5,Terasaki6}.
 Different under-structures can lead to different decay
widths, and the predictions can be compared with the experimental
measurements. For example, the value of the strong coupling constant
$g_{D_{s0}DK}$ with the assumption that the $D_{s0}(2317)$ being a
conventional scalar $c\bar s$ state is much larger than (or several
times as large as) the corresponding value with the assumption  of
being a tetraquark state \cite{WangWan06,Nielsen06}. The
$D_{s0}(2317)$ can not decay to the $D_s \gamma$ due to the angular
momentum and parity conservation, and such a final state has not
been observed; the decay $D_{s0} \to D^*_s \gamma$ is allowed and no
evidence is reported yet of the final state $D_s \gamma \gamma$
resulting from the decay chain $D_{s0} \to D^*_s \gamma \to D_s
\gamma \gamma$. The radiative decay widths of the $D_{s0} \to D_s^*
\gamma$ and $D_{s1} \to D_s \gamma$ have been calculated  with the
constituent quark model \cite{Bardeen3,Godfrey4,Li5}, the vector
meson dominance (VMD) ansatz \cite{Colangelo2} and  the light cone
QCD sum rules \cite{Colangelo1}, etc.

The amplitudes of the radiative decays $D_{s0} \to D_s^* \gamma$ and
$D_{s1} \to D_s \gamma$ can be written as
  \begin{eqnarray}
  \langle D^*_s(p, \eta) \gamma (k,\epsilon)|D_{s0}\rangle &=& e
d_S \left\{\epsilon^* \cdot \eta^* p \cdot k-\epsilon^* \cdot p
\eta^* \cdot k\right\}   \,, \nonumber \\
  \langle D_s \gamma (k,\epsilon)|D_{s1}(p, \eta)\rangle &=& ie
d_A \left\{\epsilon^* \cdot \eta p \cdot k-\epsilon^* \cdot p\eta
\cdot k  \right\}
\end{eqnarray}
due to  the Lorentz covariance.
 The $p_\mu$ and $k_\mu$ are the four momenta  of the $D_{s}^*$($D_{s1}$) and $\gamma$,
  respectively;
 the $\eta_\mu$  and  $\epsilon_\mu$ are the polarization vectors of the
  $D_{s}^*$($D_{s1}$) and $\gamma$, respectively.
 The parameters $d_S$ and $d_A$ have the dimension of inverse of the mass,
 and get contributions
from the photon couplings both to the light quark part $e_s {\bar
s}\gamma_\mu s$ and to the heavy quark part $e_c {\bar c}\gamma_\mu
c$ of the electromagnetic current, here the  $e_s$ and $e_c$ are
strange and charm quark charges in units of $e$.
 In order to determine the
amplitudes of  the $D_{s0} \to D_s^* \gamma$ and $D_{s1} \to D_s
\gamma$, we follow the VMD ansatz \cite{Colangelo2,VMD92L}. In the
heavy quark limit, the matrix elements $ \langle D^*_s(v^\prime,
\eta)|{\bar c}\gamma_\mu c|D_{s0}(v) \rangle $ and $ \langle
D_s(v^\prime)|{\bar c}\gamma_\mu c|D_{s1}(v, \eta) \rangle $ vanish
for $v\cdot v'=(M_{D_{s0}}^2+M_{D^*_s}^2)/2 M_{D_{s0}}M_{D^*_s}
\approx 1$ and $v\cdot v'=(M_{D_{s1}}^2+M_{D_s}^2)/2
M_{D_{s1}}M_{D_s} \approx 1$, here the $v_\mu$, $v'_\mu$ are the
four-velocities of the heavy mesons. We consider only the
contribution of the intermediate $\phi(1020)$,
\begin{eqnarray}
 \langle D^*_s(p, \eta) \gamma (k,\epsilon)|D_{s0}\rangle&=&
 \langle D^*_s(p,
\eta) \phi(q, \xi)|D_{s0} \rangle {i \over q^2- m_\phi^2} \langle
\gamma (k,\epsilon)|\phi(q,
\xi) \rangle \nonumber \\
&=&\langle D^*_s(p, \eta) \phi(k, \xi)|D_{s0} \rangle {i \over
k^2- m_\phi^2} f_\phi m_\phi e Q_s (-i) \epsilon^* \cdot \xi   \nonumber \\
&=&G_S(\epsilon^* \cdot \eta^* p \cdot k-\epsilon^* \cdot  p
\eta^*\cdot k)  {f_\phi \over m_\phi}   e Q_s \, , \\
 \langle D_s \gamma (k,\epsilon)|D_{s1}(p, \eta)\rangle&=&
 \langle D_s \phi(q, \xi)|D_{s1}(p,
\eta) \rangle {i \over q^2- m_\phi^2} \langle \gamma
(k,\epsilon)|\phi(q,
\xi) \rangle \nonumber \\
&=&\langle D_s \phi(k, \xi)|D_{s1} (p, \eta)\rangle {i \over
k^2- m_\phi^2} f_\phi m_\phi e Q_s (-i) \epsilon^* \cdot \xi \nonumber \\
&=&iG_A(\epsilon^* \cdot \eta p \cdot k-\epsilon^* \cdot p\eta\cdot
k)  {f_\phi \over m_\phi}   e Q_s \, ,
\end{eqnarray}
where we have use the VMD Lagrangian
\begin{eqnarray}
{\cal L}=-eQ \frac{M_\phi^2}{g_\phi}A_\alpha \phi^\alpha=-eQ f_\phi
M_\phi A_\alpha  \phi^\alpha \, ,
\end{eqnarray}
and the definitions of the strong coupling constants
\begin{eqnarray}
\langle D^*_s(p, \eta) \phi(k, \xi)|D_{s0} \rangle&=&G_S(\xi^* \cdot
\eta^*   p \cdot k-\xi^*\cdot p\eta^* \cdot  k
) \, ,\nonumber\\
\langle D_s \phi(k, \xi)|D_{s1} (p, \eta)\rangle &=&iG_A(\xi^*
\cdot\eta
 p \cdot k-\xi^*\cdot
p \eta \cdot  k )\, ,\nonumber\\
\langle 0 |{\bar s}(0)\gamma_\mu s(0)|\phi(k, \xi) \rangle &=&
f_\phi m_\phi  \xi_{ \mu} \, .
\end{eqnarray}
The $G_S$ and $G_A$ are the strong coupling constants, the $f_\phi$
is the weak decay constant of the vector meson $\phi$ and the
$\xi_\mu$ is the  polarization vector. The strong coupling constants
$G_S$ and $G_A$ can be related to the effective coupling constant
$\hat{\mu}$ in the heavy quark effective Lagrangian.
 The  Lagrange
density is set up by the hidden gauge symmetry approach with the
light vector mesons collected in a   $3 \times 3$   matrix ${\hat
V}_\mu$ \cite{VMD92L} ,
\begin{eqnarray}
 {\cal L}^\prime= i \,
\hat \mu \, Tr \left\{ {\bar S}_a H_b \sigma^{\alpha\beta}
V_{\alpha\beta}^{ba} \right\} \, ,
\end{eqnarray}
where the effective fields $H_a$ and $S_a$ stand for the  doublets
with $J^P=(0^-,1^-)$ and $(0^+,1^+)$  respectively,
\begin{eqnarray}
H_a &=& \frac{1+{\rlap{v}/}}{2}[P_{a\mu}^*\gamma^\mu-iP_a\gamma_5]
\, , \nonumber\\
S_a&=&\frac{1+{\rlap{v}/}}{2}
\left[P_{1a}^{\prime\mu}\gamma_\mu\gamma_5-P_{0a}\right] \, ,
\end{eqnarray}
where the $v_\mu$ is the four-velocity of the heavy meson and the
$a$ is a light quark flavor index. The $P_{0a}$, $P_a$, $P_a^{*}$
and $P_{1a}'$ are the scalar, pseudoscalar, vector and axial-vector
mesons respectively. The $\overline H_a= \gamma^0 H^\dagger_a
\gamma^0$, $\overline S_a= \gamma^0 S^\dagger_a \gamma^0$,
$V_{\alpha\beta}=\partial_\alpha V_\beta-\partial_\beta
V_\alpha+[V_\alpha,V_\beta]$ and $V_\alpha=i{g_V \over
\sqrt{2}}{\hat V}_\alpha$, the $g_V$ is fixed to be $g_V=5.8$ by the
KSRF rule \cite{KSRF}. Finally we obtain  the relation between the
$G_S$ ($G_A$) and $\hat{\mu}$
\begin{eqnarray}
 G_{S}&=&2\sqrt{2}\hat{\mu}g_V \,, \nonumber\\
G_{A}&=&2\sqrt{2}\hat{\mu}g_V \, .
\end{eqnarray}
The parameter $\hat{\mu}$ is a basic parameter in the heavy quark
effective Lagrangian, the precise value can lead to more deep
understanding of the relevant dynamics. It is interesting to
calculate its value with the light-cone QCD sum rules. The $\hat
\mu$ is estimated
 to $\hat \mu=-0.13 \pm 0.05 \, GeV^{-1}$ from the analysis of the $D
\to K^*$ semileptonic transitions induced by the axial weak current
\cite{VMD92L}.

 In this article, we take the point of view that
charmed mesons $D_{s0}(2317)$ and $D_{s1}(2460)$ are the
conventional $c\bar{s}$ states, calculate the values of the strong
coupling constants $G_S$ and $G_A$ ( and the corresponding
$\hat{\mu}$ ) in the framework of the light-cone QCD sum rules
approach,  and study the radiative decay widths of the $D_{s0} \to
D_s^* \gamma$ and $D_{s1} \to D_s \gamma$\footnote{The same approach
can be used to study the radiative decay widths of the $D_{s1} \to
D_s^* \gamma$, $D_{s1} \to D_{s0} \gamma$, and explore the
structures of the mesons $D_{s1}(2460)$, $D_{s0}(2317)$, that may be
our next work.}. The light-cone QCD sum rules approach carries out
the operator product expansion near the light-cone $x^2\approx 0$
instead of the short distance $x\approx 0$ while the
non-perturbative matrix elements  are parameterized by the
light-cone distribution amplitudes
 which classified according to their twists  instead of
 the vacuum condensates \cite{LCSR,LCSRreview}. The non-perturbative
 parameters in the light-cone distribution amplitudes are calculated by   the conventional QCD  sum rules
 and the  values are universal \cite{SVZ79}.

The article is arranged as: in Section 2, we derive the strong
coupling constants  $G_S$ and $G_A$ within the framework of the
light-cone QCD sum rules approach; in Section 3, the numerical
result and discussion; and in Section 4, conclusion.

\section{Strong coupling constants  $G_{S}$ and $G_{A}$ with light-cone QCD sum rules}

In the following, we write down the
 two-point correlation functions $\Pi^S_{\mu}(p,q)$ and $\Pi^A_{\mu}(p,q)$,
\begin{eqnarray}
\Pi^S_{\mu }(p,q)&=&i \int d^4x \, e^{-i q \cdot x} \,
\langle 0 |T\left\{J^S(0) J^{V}_{\mu}(x)\right\}|\phi(p)\rangle \, , \\
\Pi^A_{\mu }(p,q)&=&i \int d^4x \, e^{-i q \cdot x} \,
\langle 0 |T\left\{J^A_\mu(0) J_{5}(x)\right\}|\phi(p)\rangle \, , \\
J^S(x)&=&{\bar s}(x)   c(x)\, ,  \nonumber \\
J_5(x)&=&{\bar c}(x) i \gamma_5  s(x)\, , \nonumber \\
J^{V}_\mu(x)&=&{\bar c}(x)\gamma_\mu   s(x)\, ,\nonumber \\
J^{A}_\mu(x)&=&{\bar s}(x)\gamma_\mu \gamma_5  c(x)\, ,
\end{eqnarray}
where the currents $J^S(x)$, $J_5(x)$, $J^V_\mu(x)$ and $J^A_\mu(x)$
interpolate the mesons $D_{s0}(2317)$, $D_s$, $D^*_s$ and
$D_{s1}(2460)$, respectively, the external  state $\phi$ has the
four momentum $p_\mu$ with $p^2=m_\phi^2$. The correlation functions
$\Pi^S_{\mu}(p,q)$ and $\Pi^A_{\mu}(p,q)$ can be decomposed as
\begin{eqnarray}
\Pi^S_{\mu }(p,q)&=& \Pi_S \{\epsilon_\mu q\cdot p-p_\mu \epsilon
\cdot q \}+\cdots \, , \nonumber\\
\Pi^A_{\mu }(p,q)&=& i\Pi_A \{\epsilon_\mu (q+p)\cdot p-p_\mu
\epsilon \cdot (q+p) \}+\cdots \,
\end{eqnarray}
due to the Lorentz covariance, here the $\epsilon_\mu$ is the
polarization vector of the $\phi$ meson.

According to the basic assumption of current-hadron duality in the
QCD sum rules approach \cite{SVZ79}, we can insert  a complete
series of intermediate states with the same quantum numbers as the
current operators $J^S(x)$, $J_5(x)$, $J^V_\mu(x)$ and $J^A_\mu(x)$
into the correlation functions $\Pi^S_{\mu}(p,q)$ and
$\Pi^A_{\mu}(p)$ to obtain the hadronic representation. After
isolating the ground state contributions from the pole terms of the
mesons $D_{s0}(2317)$, $D_s$, $D^*_s$ and $D_{s1}(2460)$, we get the
following results,
\begin{eqnarray}
\Pi^S_{\mu }(p,q)&=&\frac{\langle0| J^S(0)|
D_{s0}(q+p)\rangle\langle D_{s0}| D^*_s \phi \rangle  \langle
D^*_s(q)|J^V_\mu(0)| 0\rangle}
  {\left\{M_{D_{s0}}^2-(q+p)^2\right\}\left\{M_{D^*_s}^2-q^2\right\}}  + \cdots \nonumber \\
&=&\frac{f_{D_{s0}}f_{D^*_s}M_{D_{s0}}M_{D^*_s}G_S}
  {\left\{M_{D_{s0}}^2-(q+p)^2\right\}\left\{M_{D^*_s}^2-q^2\right\}}\{\epsilon_\mu q\cdot p-p_\mu \epsilon
\cdot q \}  + \cdots  , \\
\Pi^A_{\mu }(p,q)&=&\frac{\langle0| J^A_\mu(0)|
D_{s1}(q+p)\rangle\langle D_{s1}| D_s \phi \rangle  \langle
D_s(q)|J_5(0)| 0\rangle}
  {\left\{M_{D_{s1}}^2-(q+p)^2\right\}\left\{M_{D_s}^2-q^2\right\}}  + \cdots \nonumber \\
&=&i\frac{f_{D_{s1}}f_{D_s}M_{D_{s1}}M_{D_s}^2G_A}
  {(m_c+m_s)\left\{M_{D_{s1}}^2-(q+p)^2\right\}\left\{M_{D_s}^2-q^2\right\}}\{\epsilon_\mu (q+p)\cdot p-p_\mu
\epsilon \cdot (q+p) \}  \nonumber\\
&&+ \cdots  ,
\end{eqnarray}
where the following definitions have been used,
\begin{eqnarray}
\langle0 | J^S(0)|D_{s0}(p)\rangle&=&f_{D_{s0}}M_{D_{s0}}\,, \nonumber\\
\langle0 | {J_5}^+(0)|D_{s}(p)\rangle&=&\frac{f_{D_{s}}M_{D_{s}}^2}{m_c+m_s}\,, \nonumber\\
\langle0 |
{J^V_{\mu}}^+(0)|D^*_{s}(p,\eta)\rangle&=&f_{D^*_{s}}M_{D^*_{s}}\eta_\mu\,
,\nonumber \\
\langle0 |
J^A_{\mu}(0)|D_{s1}(p,\eta)\rangle&=&f_{D_{s1}}M_{D_{s1}}\eta_\mu\,
,\nonumber
\end{eqnarray}
where the $f_{D_{s0}}$, $f_{D_s}$, $f_{D^*_s}$ and $f_{D_{s1}}$ are
the weak decay constants  of the $D_{s0}(2317)$, $D_s$, $D^*_s$ and
$D_{s1}(2460)$, respectively.
 In Eqs.(13-14), we have not shown the contributions from the high
resonances and continuum states explicitly as they are suppressed
due to the double Borel transformation.

In the following, we briefly outline the  operator product expansion
for the correlation functions $\Pi^S_{\mu }(p,q)$ and $\Pi^A_{\mu
}(p,q)$ in perturbative QCD theory. The calculations are performed
at the large space-like momentum regions $(q+p)^2\ll 0$  and $q^2\ll
0$, which correspond to the small light-cone distance $x^2\approx 0$
required by the validity of the operator product expansion approach.
We write down the propagator of a massive quark in the external
gluon field in the Fock-Schwinger gauge firstly \cite{Belyaev94},
\begin{eqnarray}
\langle 0 | T \{q_i(x_1)\, \bar{q}_j(x_2)\}| 0 \rangle &=&
 i \int\frac{d^4k}{(2\pi)^4}e^{-ik(x_1-x_2)}
\frac{\not\!k +m}{k^2-m^2} +\cdots\, ,
\end{eqnarray}
here we have neglected the contributions from the gluons $G_{\mu \nu
}$. The contributions proportional to the $G_{\mu\nu}$ can give rise
to three-particle (and four-particle) meson distribution amplitudes
with a gluon (or quark-antiquark pair) in addition to the two
valence quarks, their corrections are usually not expected to play
any significant roles\footnote{For examples, in the decay $B \to
\chi_{c0}K$, the factorizable contribution is zero and the
non-factorizable contributions from the soft hadronic matrix
elements are too small to accommodate the experimental data
\cite{WangLH}; the net contributions from the three-valence particle
light-cone distribution amplitudes to the strong coupling constant
$g_{D_{s1}D^*K}$ are rather small, about $20\%$ \cite{Wang0611}. The
contributions of the three-particle (quark-antiquark-gluon)
distribution amplitudes of the mesons are always of minor importance
comparing with the two-particle (quark-antiquark) distribution
amplitudes in the light-cone QCD sum rules.   In our previous work,
we study the four form-factors $f_1(Q^2)$, $f_2(Q^2)$, $g_1(Q^2)$
and $g_2(Q^2)$ of the $\Sigma \to n$ in the framework of the
light-cone QCD sum rules approach up to twist-6 three-quark
light-cone distribution amplitudes and obtain satisfactory results
\cite{Wang06}. In the light-cone QCD sum rules,
 we can neglect the contributions from the valence gluons and make relatively rough estimations.}. Substituting the above $c$ quark
propagator and the corresponding $\phi$ meson light-cone
distribution amplitudes into the correlation functions
$\Pi^S_{\mu}(p,q)$ and $\Pi^A_{\mu}(p,q)$ in Eqs.(9-10) and
completing the integrals over the variables  $x$ and $k$, finally we
obtain the results,
\begin{eqnarray}
\Pi_S &=&-f_\phi^T  \int_0^1 du
\frac{\phi_\perp(u)}{AA}+\left[f_\phi^T-f_\phi\frac{2m_s}{m_\phi}\right]m_\phi^2
\int_0^1 duu \frac{h_{||}^{(s)}(u)
}{AA^2} \nonumber\\
&&+\frac{f_\phi^T m_\phi^2}{4} \int_0^1 du A_\perp(u) \left[
\frac{1}{AA^2} +\frac{2m_c^2}{AA^3}\right]
+2f_\phi^T m_\phi^2 \int_0^1 du \int_0^u d\tau \int_0^\tau dt \frac{B_\perp (t)}{AA^2}\nonumber \\
&&-f_\phi^T m_\phi^2 \int_0^1 du u \int_0^u dt C_\perp(t)\left[
\frac{1}{AA^2} +\frac{2m_c^2}{AA^3}\right]-2f_\phi m_\phi m_c
 \int_0^1 du u \frac{g_{\perp}^{(v)}}{AA^2} \nonumber \\
 &&+2f_\phi m_\phi^3 m_c \int_0^1 du u \int_0^u d \tau \int_0^\tau
 dt \frac{C(t)}{AA^3} \, ,
\end{eqnarray}
\begin{eqnarray}
\Pi_A &=&-f_\phi^T  \int_0^1 du
\frac{\phi_\perp(u)}{AA}-\left[f_\phi^T-f_\phi\frac{2m_s}{m_\phi}\right]m_\phi^2
\int_0^1 du u\frac{h_{||}^{(s)}(u)
}{AA^2} \nonumber\\
&&+\frac{f_\phi^T m_\phi^2}{4} \int_0^1 du A_\perp(u) \left[
\frac{1}{AA^2} +\frac{2m_c^2}{AA^3}\right]
+2f_\phi^T m_\phi^2 \int_0^1 du \int_0^u d\tau \int_0^\tau dt \frac{B_\perp (t)}{AA^2}\nonumber \\
&&-f_\phi^T m_\phi^2 \int_0^1 du u \int_0^u dt C_\perp(t)\left[
\frac{1}{AA^2} +\frac{2m_c^2}{AA^3}\right]-2f_\phi m_\phi m_c
 \int_0^1 du u \frac{g_{\perp}^{(v)}}{AA^2} \nonumber \\
 &&+2f_\phi m_\phi^3 m_c \int_0^1 du u \int_0^u d \tau \int_0^\tau
 dt \frac{C(t)}{AA^3} \, ,
\end{eqnarray}
where
\begin{eqnarray}
AA&=&m_c^2-(q+up)^2 \, .\nonumber
\end{eqnarray}
In calculation, the  two-particle  $\phi$ meson light-cone
distribution amplitudes have been used \cite{VMLC}, the explicitly
expressions are given in the appendix. The parameters in the
light-cone distribution amplitudes are scale dependent and can be
estimated with the QCD sum rules approach \cite{VMLC}. In this
article, the energy scale $\mu$ is chosen to be  $\mu=1GeV$.

Now we perform the double Borel transformation with respect to  the
variables $Q_1^2=-(p+q)^2$  and  $Q_2^2=-q^2$ for the correlation
functions $\Pi_S$ and $\Pi_A$ in Eqs.(13-14),   and obtain the
analytical expressions of the invariant functions in the hadronic
representation,
\begin{eqnarray}
B_{M_2^2}B_{M_1^2}\Pi_S&=&\frac{
G_Sf_{D^*_s}f_{D_{s0}}M_{D^*_s}M_{D_{s0}}}{M_1^2M_2^2}
\exp\left[-\frac{M^2_{D_{s0}}}{M_1^2}
-\frac{M^2_{D^*_s}}{M_2^2}\right] +\cdots, \\
B_{M_2^2}B_{M_1^2}\Pi_A&=&\frac{
G_Af_{D_s}f_{D_{s1}}M_{D_s}^2M_{D_{s1}}}{M_1^2M_2^2(m_c+m_s)}
\exp\left[-\frac{M^2_{D_{s1}}}{M_1^2}
-\frac{M^2_{D_s}}{M_2^2}\right] +\cdots,
\end{eqnarray}
here we have not shown  the contributions from the high resonances
and continuum states  explicitly for simplicity. In order to match
the duality regions below the thresholds $s_0$ and $s_0'$ for the
interpolating currents $J^{S}(x)$ (or $J^A_\mu(x)$)  and
$J^V_\mu(x)$(or $J_{5}(x)$) respectively, we can express the
correlation functions $\Pi$ (denote the $\Pi_S$ and $\Pi_A$) at the
level of quark-gluon degrees of freedom into the following form,
\begin{eqnarray}
\Pi&=& \int ds \int ds' \frac{\rho(s,s')}{
\left\{s-(q+p)^2\right\}\left\{s'-q^2\right\}} \, ,
\end{eqnarray}
where the $\rho(s,s')$ are spectral densities, then perform the
double Borel transformation with respect to the variables $Q_1^2$
and $Q_2^2$ directly. However, the analytical expressions of the
spectral densities $\rho(s,s')$ are hard to obtain, we have to
resort to some approximations.  As the contributions
 from the higher twist terms  are  suppressed by more powers of
 $\frac{1}{m_c^2-(q+up)^2}$, the net contributions of the  twist-3 and twist-4
  terms are of minor
importance, less  than  $20\%$, the continuum subtractions will not
affect the results remarkably. The dominating contribution comes
from the two-particle twist-2 term involving the $\phi_\perp(u)$. We
preform the same trick as Refs.\cite{Belyaev94,Kim} and expand the
amplitude $\phi_\perp(u)$ in terms of polynomials of $1-u$,
\begin{eqnarray}
\phi_\perp(u)=\sum_{k=0}^N b_k(1-u)^k=\sum_{k=0}^N b_k
\left(\frac{s-m_c^2}{s-q^2}\right)^k,
\end{eqnarray}
then introduce the variable $s'$ and the spectral density is
obtained.

After straightforward calculations, we obtain the final expressions
of the double Borel transformed correlation functions $\Pi
(M_1^2,M_2^2)$ at the level of quark-gluon degrees of freedom. The
masses of  the charmed mesons are $M_{D_{s1}}=2.46GeV$,
$M_{D_{s0}}=2.317GeV$, $M_{D_{s}}=1.97GeV$ and $M_{D^*_s}=2.112GeV$,
$\frac{M_{D^*_s}}{M_{D^*_s}+M_{D_{s0}}}\approx0.48$,
$\frac{M_{D_s}}{M_{D_s}+M_{D_{s1}}}\approx0.45$,
 there exists an overlapping working window for the two Borel
parameters $M_1^2$ and $M_2^2$, it's convenient to take the value
$M_1^2=M_2^2$. We introduce the threshold parameter $s_0$ and make
the simple replacement,
\begin{eqnarray}
e^{-\frac{m_c^2+u_0(1-u_0)m_\phi^2}{M^2}} \rightarrow
e^{-\frac{m_c^2+u_0(1-u_0)m_\phi^2}{M^2} }-e^{-\frac{s_0}{M^2}}
\nonumber
\end{eqnarray}
 to subtract the contributions from the high resonances  and
  continuum states \cite{Belyaev94}, finally we obtain the sum rules for the strong coupling
  constants $G_S$ and $G_A$,
\begin{eqnarray}
&&G_Sf_{D_{s0}}f_{D^*_s}M_{D_{s0}}M_{D^*_s}\exp\left\{-\frac{M_{D_{s0}}^2}{M^2_1}-\frac{M^2_{D^*_s}}{M_2^2}\right\}\nonumber\\
&=& -f_\phi^TM^2\phi_\perp(u_0)\left\{\exp\left[- \frac{BB}{M^2}
\right]-\exp\left[- \frac{s_0}{M^2} \right] \right\} \nonumber\\
&&+\exp\left[- \frac{BB}{M^2} \right]\left\{
\left[f_\phi^T-f_\phi\frac{2m_s}{m_\phi}\right]m_\phi^2
 h_{||}^{(s)}(u_0)
u_0 +\frac{f_\phi^T m_\phi^2A_\perp(u_0)}{4}  \left[1
+\frac{m_c^2}{M^2}\right]\right.\nonumber\\
&&+2f_\phi^T m_\phi^2  \int_0^{u_0} d\tau \int_0^\tau dt B_\perp (t)
-f_\phi^T m_\phi^2  u_0 \int_0^{u_0} dt C_\perp(t)\left[1
+\frac{m_c^2}{M^2}\right]\nonumber \\
&&\left.-2f_\phi m_\phi m_c  u_0g_{\perp}^{(v)}(u_0)
 +f_\phi m_\phi^3 m_c  u_0 \int_0^{u_0} d \tau \int_0^\tau
 \frac{C(t)}{M^2} \right\} \, ,
\end{eqnarray}
\begin{eqnarray}
&&G_Af_{D_{s1}}f_{D_s}M_{D_{s1}}\frac{M_{D_s}^2}{m_c+m_s}\exp\left\{-\frac{M_{D_{s1}}^2}{M^2_1}-
\frac{M^2_{D_s}}{M_2^2}\right\}\nonumber\\
&=& -f_\phi^TM^2\phi_\perp(u_0)\left\{\exp\left[- \frac{BB}{M^2}
\right]-\exp\left[- \frac{s_0}{M^2} \right] \right\}\nonumber \\
&&+\exp\left[- \frac{BB}{M^2} \right]\left\{
-\left[f_\phi^T-f_\phi\frac{2m_s}{m_\phi}\right]m_\phi^2
 h_{||}^{(s)}(u_0)
u_0 +\frac{f_\phi^T m_\phi^2A_\perp(u_0)}{4}  \left[1
+\frac{m_c^2}{M^2}\right]\right.\nonumber\\
&&+2f_\phi^T m_\phi^2  \int_0^{u_0} d\tau \int_0^\tau dt B_\perp (t)
-f_\phi^T m_\phi^2  u_0 \int_0^{u_0} dt C_\perp(t)\left[1
+\frac{m_c^2}{M^2}\right]\nonumber \\
&&\left.-2f_\phi m_\phi m_c  u_0g_{\perp}^{(v)}(u_0)
 +f_\phi m_\phi^3 m_c  u_0 \int_0^{u_0} d \tau \int_0^\tau
 \frac{C(t)}{M^2} \right\} \, ,
\end{eqnarray}
where
\begin{eqnarray}
BB&=&m_c^2+u_0(1-u_0)m_\phi^2 \, ,\nonumber \\
u_0&=&\frac{M_1^2}{M_1^2+M_2^2}\, , \nonumber \\
M^2&=&\frac{M_1^2M_2^2}{M_1^2+M_2^2} \, .
\end{eqnarray}

\section{Numerical result and discussion}
The parameters are taken as $m_s=(140\pm 10 )MeV$, $m_c=(1.25\pm
0.10)GeV$, $f_\phi=(0.215\pm0.005)GeV$,
$f_\phi^{\perp}=(0.186\pm0.009)GeV$, $a_2^{\perp}=0.2\pm0.2$,
$a_2^{\parallel}=0.2\pm0.1$, $\varsigma_3=0.032\pm 0.010$,
$\varsigma_4=0.15\pm 0.10$, $\varsigma_4^T=0.10\pm 0.05$,
$\widetilde{\varsigma}_4^T=-0.10\pm 0.05$, $\omega_3^A=-2.1\pm1.0$,
$\omega_3^V=3.8\pm1.8$, $\omega_3^T=7.0\pm7.0$,
 $m_\phi=1.02GeV$, $M_{D_{s0}}=2.317GeV$, $M_{D_{s1}}=2.46GeV$,
$M_{D^*_s}=2.112GeV$, $M_{D_s}=1.97GeV$,
$f_{D^*_s}=(0.26\pm0.02)GeV$, $f_{D_s}=(0.26\pm0.02)GeV$,
$f_{D_{s0}}=(0.225\pm0.025)GeV$ and $f_{D_{s1}}=(0.225\pm0.025)GeV$
\cite{Colangelo2,decayC}. The duality thresholds $s_0$ in
Eqs.(22-23) are taken as $s_0=(7.0-8.0)GeV^2$
($\sqrt{s_0}=(2.6-2.8)GeV$)
 to avoid possible  contaminations from the high resonances and
continuum states, in this region, the numerical results are not
sensitive to the threshold  parameters $s_0$. The Borel parameters
are chosen as $ M_1^2=M_2^2$ and $M^2=(3.5-8)GeV^2$, in those
regions, the values of  the strong coupling constants $G_S$ and
$G_A$ are rather stable from the sum rules in Eqs.(22-23) with the
simple subtraction.

The uncertainties of the eleven parameters $f_\phi$,
$f_\phi^{\perp}$, $a_2^{\parallel}$, $\omega_3^T$, $\omega_3^V$,
$\omega_3^A$, $\varsigma_3$, $\varsigma_4$, $\varsigma_4^T$,
$\widetilde{\varsigma}_4^T$   and $m_s$ can not result in large
uncertainties for the numerical values. The main uncertainties come
from the six parameters $f_{D_{s0}}$,
 $f_{D_{s1}}$, $f_{D^*_s}$, $f_{D_s}$, $a_2^{\perp}$
and  $m_c$,  the variations of those parameters can lead to
relatively large changes for the numerical values, while the
dominating uncertainty comes from the $a_2^{\perp}$, which are shown
in the Fig.1, refining the parameter $a_2^{\perp}$ is of great
importance. Taking into account all the uncertainties, finally we
obtain the numerical results of the strong coupling constants $G_S$
and $G_A$, which are shown in the Fig.2,
\begin{eqnarray}
  G_S &=&-(4.4\pm2.0) \frac{1}{GeV} \, ,\nonumber \\
G_A &=&-(3.9\pm1.6) \frac{1}{GeV}   \, , \\
\hat{\mu}_S &=&-(0.27\pm0.12) \frac{1}{GeV} \, , \nonumber\\
\hat{\mu}_A &=&-(0.24\pm0.10) \frac{1}{GeV}   \, .
\end{eqnarray}

\begin{figure}
\centering
  \includegraphics[totalheight=7cm,width=7cm]{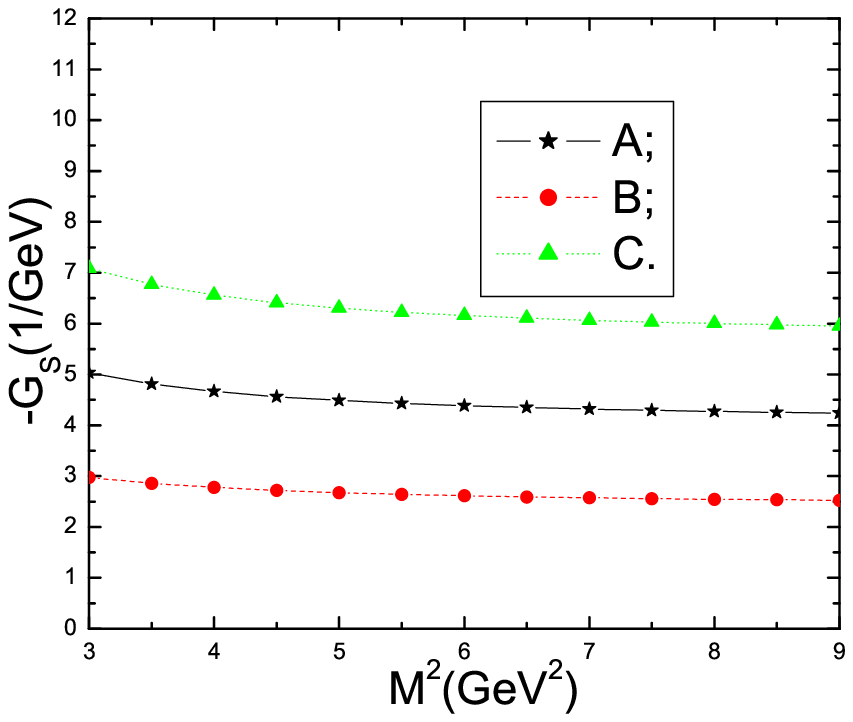}
  \includegraphics[totalheight=7cm,width=7cm]{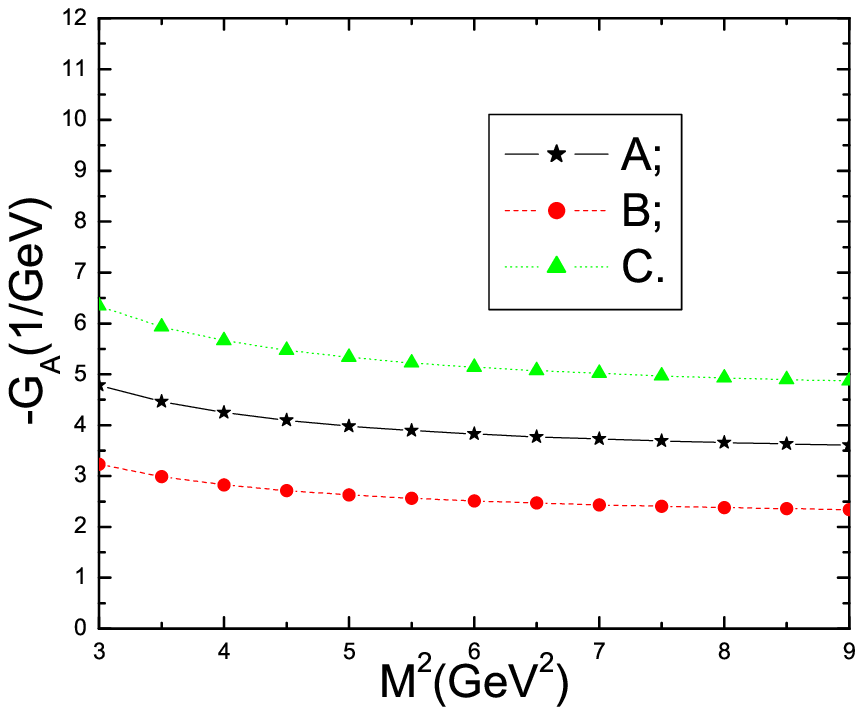}
   \caption{The   $G_S$ and $G_A$ with the Borel parameter $M^2$ for $a_2^{\perp}=0.2$ ($A$), $a_2^{\perp}=0.4$ ($B$) and
   $a_2^{\perp}=0.0$ ($C$). }
\end{figure}
\begin{figure}
\centering
  \includegraphics[totalheight=7cm,width=7cm]{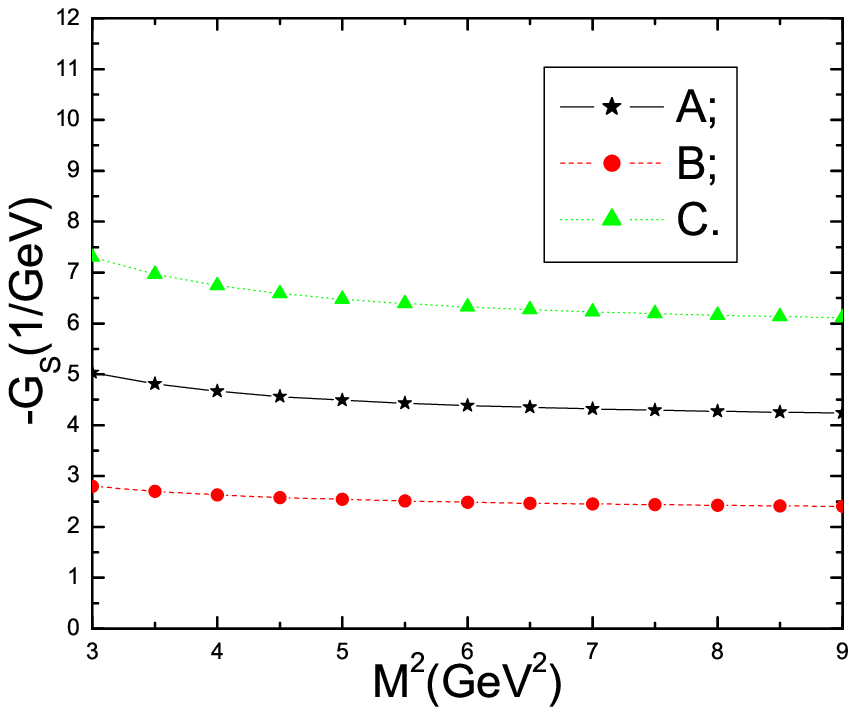}
  \includegraphics[totalheight=7cm,width=7cm]{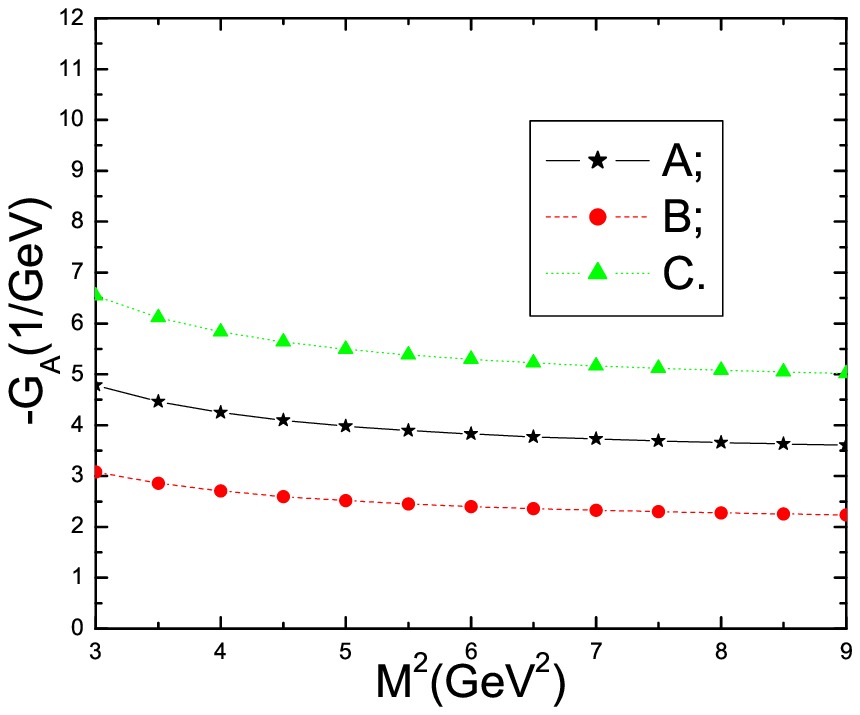}
   \caption{The   $G_S$ and $G_A$ with the Borel parameter $M^2$, the $A$, $B$ and $C$ stand for the central value,
   down limit and up limit, respectively. }
\end{figure}
 The
corresponding values of the parameter $\hat{\mu}$ are larger than
the existing estimation $\hat \mu=-0.13 \pm 0.05 \, GeV^{-1}$ from
the analysis of the $D \to K^*$ semileptonic transitions induced by
the axial weak current \cite{VMD92L}. From the numerical values of
the strong coupling constants $G_S$ and $G_A$, we can obtain the
decay widths,
\begin{eqnarray}
  \Gamma(D_{s0} \to D^*_s \gamma) &=&(1.3-9.9) KeV \, , \nonumber \\
\Gamma(D_{s1} \to D_s \gamma) &=&(5.5-31.2) KeV \,    .
\end{eqnarray}
The comparison  with the results from other approaches is presented
in the Table.1.
\begin{table}[htb]
\begin{center}
\begin{tabular}{c|c|c|c|c|c|c|c} \hline
KeV
&\cite{Colangelo1}&\cite{Colangelo2}&\cite{Bardeen3}&\cite{Godfrey4}&
\cite{Li5}&\cite{Terasaki6} &This work \\\hline
 $\Gamma(D_{s0}\to D^*_{s}\gamma)$& $ 4-6$ &$ 1$&$1.74$&$1.9$&$1.1$ &35&1.3-9.9  \\\hline
 $\Gamma(D_{s1}\to D_{s}\gamma)$&$19-29$ &
 &$5.08$&$6.2$&$0.6-2.9$& &5.5-31.2
\\\hline
\end{tabular}
\end{center}
\caption{The decay widths of the $D_{s0}\to D^*_{s}\gamma$,
$D_{s1}\to D_{s}\gamma$. }
\end{table}

\section{Conclusion}
In this article, we take the point of view that the  charmed
 mesons $D_{s0}(2317)$ and  $D_{s1}(2460)$ with the spin-parity $0^+$ and $1^+$ respectively are
  the conventional $c\bar{s}$
states and  calculate the strong coupling constants $G_S$ and $G_A$
in the framework of the light-cone QCD sum rules approach. The
strong coupling constants $G_S$ and $G_A$ are related to the basic
parameter $\hat{\mu}$ in the heavy quark effective Lagrangian, the
numerical value of the $\hat{\mu}$ is larger than the existing
estimation. With the assumption of the vector meson dominance of the
intermediate $\phi$, we study the radiative decays $D_{s0}\to D_s^*
\gamma $ and $D_{s1}\to D_s \gamma $, the numerical values of the
decay widths  are compatible with the existing estimations, further
experimental data
 can conform or reject the  assumption of
the two-quark substructure. Just like the scalar mesons $f_0(980)$
and $a_0(980)$, the scalar meson $D_{s0}(2317)$ and the axial-vector
meson $D_{s1}(2460)$ may have small $c\bar{s}$ kernels of typical
$c\bar{s}$ meson size. The strong couplings  to virtual intermediate
hadronic states (or the virtual mesons loops) can result in  smaller
masses  than the conventional
 $0^+$ and $1^+$ mesons in the constituent quark models, enrich the
pure $c\bar{s}$ states with other components. The $D_{s0}(2317)$ and
$D_{s1}(2460)$ may spend part (or most part) of their lifetimes as
virtual $ D K $ and $ D^* K $ states, respectively.
 \section*{Appendix}
 The light-cone distribution amplitudes of the $\phi$ meson are defined
 by
\begin{eqnarray}
\langle 0| {\bar s} (0) \gamma_\mu s(x) |\phi(p)\rangle& =& p_\mu
f_\phi m_\phi \frac{\epsilon \cdot x}{p \cdot x} \int_0^1 du  e^{-i
u p\cdot x} \left\{\phi_{\parallel}(u)+\frac{m_\phi^2x^2}{16}
A(u)\right\}\nonumber\\
&&+\left[ \epsilon_\mu-p_\mu \frac{\epsilon \cdot x}{p \cdot x}
\right]f_\phi m_\phi
\int_0^1 du  e^{-i u p \cdot x} g_{\perp}^{(v)}(u)  \nonumber\\
&&-\frac{1}{2}x_\mu \frac{\epsilon \cdot x}{(p \cdot x)^2} f_\phi m_\phi^3 \int_0^1 du e^{-iup \cdot x}C(u) \, ,\nonumber\\
 \langle 0| {\bar s} (0)  s(x) |\phi(p)\rangle  &=& \frac{i}{
2}\left[f_\phi^T-f_\phi \frac{2m_s}{m_\phi}\right]m_\phi^2\epsilon \cdot x  \int_0^1 du  e^{-i u p \cdot x} h_{\parallel}^{(s)}(u)  \, ,  \nonumber\\
\langle 0| {\bar s} (0) \sigma_{\mu \nu}  s(x) |\phi(p)\rangle
&=&i[\epsilon_\mu p_\nu-\epsilon_\nu p_\mu] f_\phi^T \int_0^1 du
e^{-i u p \cdot x}  \left\{\phi_{\perp}(u)+\frac{m_\phi^2x^2}{16}
A_{\perp}(u) \right\}   \nonumber\\
&&+i[p_\mu x_\nu-p_\nu x_\mu] f_\phi^T m_\phi^2\frac{\epsilon \cdot
x}{(p \cdot x)^2} \int_0^1 du e^{-i u p \cdot x}
 B_{\perp}(u)   \nonumber\\
 &&+i\frac{1}{2}[\epsilon_\mu x_\nu-\epsilon_\nu x_\mu] f_\phi^T m_\phi^2\frac{1}{p
\cdot x} \int_0^1 du e^{-i u p \cdot x}
 C_{\perp}(u)   \, .
\end{eqnarray}
The  light-cone distribution amplitudes are parameterized as
\begin{eqnarray}
\phi_{\parallel}(u,\mu)&=&6u(1-u) \left\{1+a_2^{\parallel}
\frac{3}{2}(5\xi^2-1) \right\}\, , \nonumber\\
\phi_{\perp}(u,\mu)&=&6u(1-u) \left\{1+a_2^{\perp}
\frac{3}{2}(5\xi^2-1) \right\}\, , \nonumber\\
g_{\perp}^{(v)}(u,\mu)&=&\frac{3}{4}(1+\xi^2)+\left\{ \frac{3}{7}a_2^{\parallel}+ 5\varsigma_3\right\}(3\xi^2-1)\nonumber \\
&&+\left\{\frac{9}{112}a_2^{\parallel}+\frac{15}{64}\varsigma_3(3\omega^V_3-\omega^A_3 )\right\}(3-30\xi^2+35\xi^4)\, ,  \nonumber \\
g_3(u,\mu)&=&1+\left\{ -1-\frac{2}{7}a_2^{\parallel}+\frac{40}{3}\varsigma_3 -\frac{20}{3}\varsigma_4\right\}C_2^{\frac{1}{2}}(\xi)\nonumber \\
&&+\left\{-\frac{27}{28}a_2^{\parallel} +\frac{5}{4}\varsigma_3 -\frac{15}{16}\varsigma_3(\omega^A_3+3\omega^V_3)\right\}C_4^{\frac{1}{2}}(\xi)\, ,  \nonumber \\
h_3(u,\mu)&=&1+\left\{ -1+\frac{3}{7}a_2^{\perp}-10(\varsigma_4^T+\widetilde{\varsigma}_4^T)\right\}C_2^{\frac{1}{2}}(\xi)\nonumber \\
&&+\left\{-\frac{3}{7}a_2^{\perp} -\frac{15}{8}\varsigma_3\omega_3^T\right\}C_4^{\frac{1}{2}}(\xi)\, ,  \nonumber \\
h_{\parallel}^{(s)}(u,\mu)&=&6u(1-u)
\left\{1+(\frac{1}{4}a_2^{\parallel}+\frac{5}{8}\varsigma_3\omega_3^T)
(5\xi^2-1) \right\}\, , \nonumber\\
h_{\perp}^{(s)}(u,\mu)&=&3\xi^2+\frac{3}{2}a_2^{\perp}\xi^2(5\xi^2-3)+\frac{15}{16}\varsigma_3 \omega_3^T(3-30\xi^2+35\xi^4) , \nonumber\\
A_{\perp}(u,\mu)&=&30u^2(1-u)^2\left\{\frac{2}{5}
+\frac{4}{35}a_2^{\perp}+\frac{4}{3}\varsigma_4^T-\frac{8}{3}\widetilde{\varsigma}_4^T\right\}\,
, \nonumber \\
C(u,\mu)&=&g_3(u,\mu)+\phi_{\parallel}(u,\mu)-2g_{\perp}^{(v)}(u,\mu)
\, ,\nonumber \\
B_{\perp}(u,\mu)&=&h_{\perp}^{(s)}(u,\mu)-\frac{1}{2}\phi_{\perp}(u,\mu)-\frac{1}{2}h_3(u,\mu)\, ,\nonumber \\
C_{\perp}(u,\mu)&=&h_3(u,\mu)-\phi_{\perp}(u,\mu) \, ,
\end{eqnarray}
where the $\xi=2u-1$, and  the   $ C_2^{\frac{1}{2}}$ and
$C_4^{\frac{1}{2}}$
   are Gegenbauer polynomials \cite{VMLC}.
\section*{Acknowledgments}
This  work is supported by National Natural Science Foundation,
Grant Number 10405009,  and Key Program Foundation of NCEPU.

\end{document}